\documentclass[reprint,superscriptaddress]{revtex4-1}

\usepackage[OT2,T1]{fontenc}

\newcommand\textcyr[1]{{\fontencoding{OT2}\selectfont #1}}

\usepackage{hyperref}
\usepackage{amsmath} % for \text command

\usepackage{graphicx} %includegraphics
\graphicspath{ {./fig/} }

\usepackage{color}
\newcommand{\M}[1]{\mathcal{#1}}
\newcommand{\abs}[1]{\left\lvert #1 \right\rvert}

\begin{document}

\title{Multistability of a chiral semiconductor microcavity: a self-consistent approach}

\author{O. A. Dmitrieva}
\affiliation{Skolkovo Institute of Science and Technology, 1 21205 Moscow, Russia}
\affiliation{Lomonosov Moscow State University, 119991 Moscow, Russia}
\affiliation{Prokhorov Institute of General Physics, Russian Academy of Sciences, 119991 Moscow, Russia}

\author{N. A. Gippius}
\affiliation{Skolkovo Institute of Science and Technology, 1 21205 Moscow, Russia}

\author{S. G. Tikhodeev}
\affiliation{Skolkovo Institute of Science and Technology, 1 21205 Moscow, Russia}
\affiliation{Lomonosov Moscow State University, 119991 Moscow, Russia}
\affiliation{Prokhorov Institute of General Physics, Russian Academy of Sciences, 119991 Moscow, Russia}

%{\it Keywords}: semiconductor Bragg microcavity, exciton-polaritons, nonlinearity, bistability, multistability, photonic crystal, chirality

\begin{abstract}
	We calculate the effects of polariton bi- and multistability in a semiconductor Bragg microcavity with multiple
	quantum wells and a chiral photonic crystal on the upper mirror for resonant coherent pumping normal to the structure.
	Even if the system is not optimized for obtaining photoluminescence with a high degree of circular polarization in
	the spontaneous mode, it is shown that linear-polarized pumping can cause nonlinear switching to states with a degree of
	circular polarization of polaritons up to 90\%. Calculations were performed in both the mean-field and self-consistent
	approximations, accounting for the difference in exciton density among the microcavity's quantum wells.
    This paper in Russian is published as O. A. Dmitrieva, et al., Zh. Exp. Teor. Fiz. {\bf 169}, 181 (2026)~\cite{Dmitrieva2026}.
\end{abstract}

\maketitle

\section{Introduction}

In recent years, the properties of compact sources of circularly polarised light based on chiral heterostructures, including lasers,  have attracted considerable attention from researchers~\cite{Konishi2011,Maksimov2014,Lobanov2015,Demenev2016,Tanaka2020,Qu2021,Gao2021,Fong2021,Zhang2022,Maksimov2022,Tsai2024,Toftul2024,Takahashi2025,Gromyko2025,Valenko2025rus}.
The reason for this is their relevance in spectroscopy, sensorics and information technology.
One implementation of such systems is a diode semiconductor laser with highly circularly polarized radiation, up to 90\%~\cite{Maksimov2022,Maksimov2022arus,Valenko2025rus}, based on a semiconductor Bragg microcavity with exciton-polaritons in the active layer's quantum wells and a chiral modulated upper mirror.  Due to the high nonlinearity associated with cavity polaritons, semiconductor microcavities exhibit strong bistability and multistability effects under resonant optical pumping.~\cite{Baas2004,Gippius2004,Gippius2007,Sarkar2010,Gavrilov2013,Brichkin2015,Gavrilov2020rus}.
The characteristic times of bi- and multistable switching appear to be in the picosecond range~\cite{Gippius2004,Gavrilov2020rus}, making them potentially interesting for practical applications.

\begin{figure}
    \begin{center}
        \begin{minipage}[h]{0.09\textwidth}
        {(a) \hspace*{0.01\textwidth}
            \includegraphics[width = \textwidth]{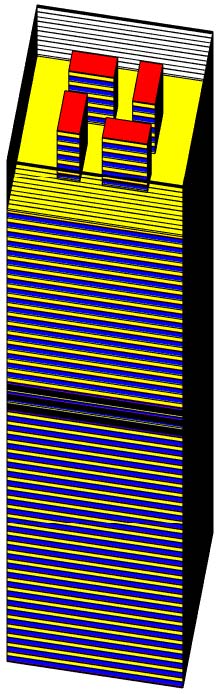}
            }
        \end{minipage}
        %\hfill
        \begin{minipage}[h]{0.15\textwidth}
            (b) \hspace*{-0.01\textwidth}
            \includegraphics[width = \textwidth]{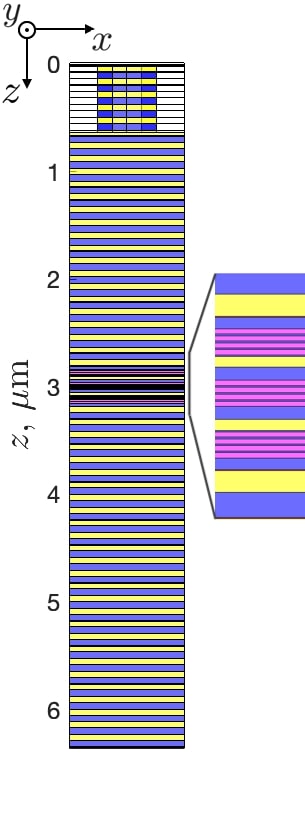}
        \end{minipage}
        %\hfill
        \hspace*{0.01\textwidth}
        \begin{minipage}[h]{0.21\textwidth}
             (c) \includegraphics[width = \textwidth]{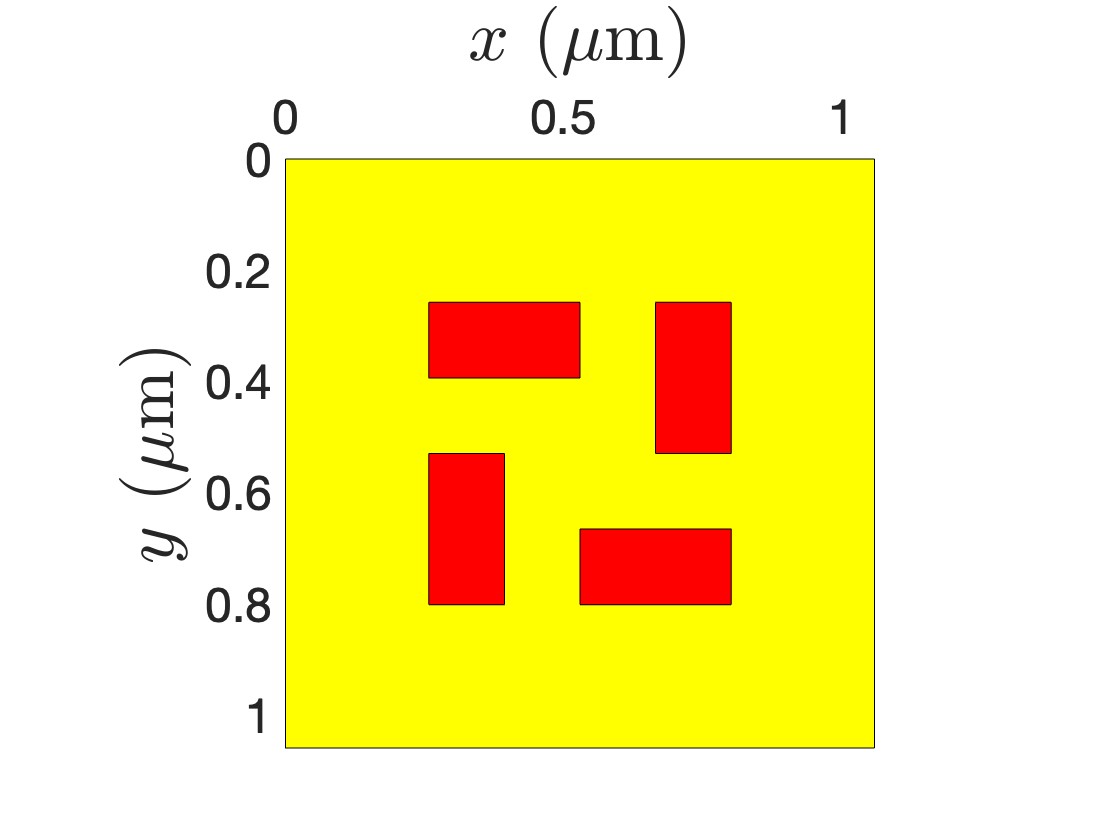}
            (d) \includegraphics[width = \textwidth]{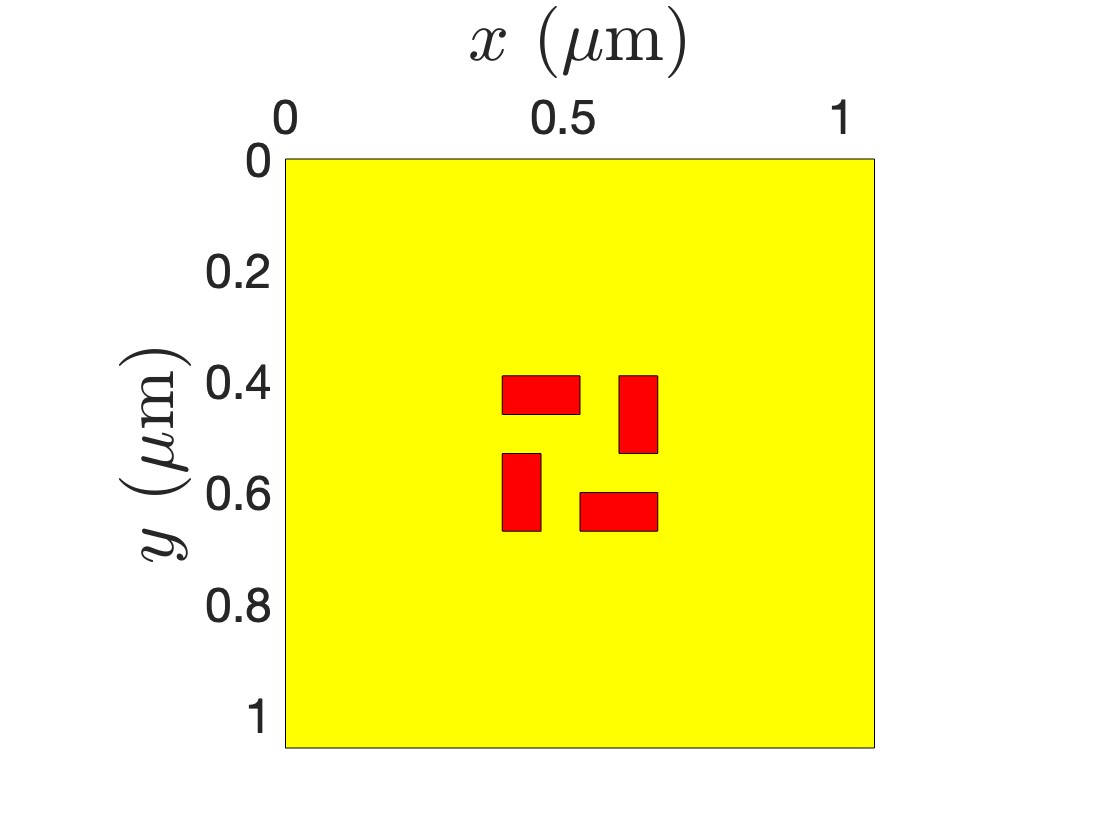}
        \end{minipage}
    \end{center}
    \caption{Schematic representation of a chiral semiconductor microcavity. (a) Elementary cell of the structure. AlGaAs layers are shown in yellow, AlAs in blue, GaAs in red in the cap layer region, and magenta in quantum wells.
    (b) and (c,d) Vertical and horizontal cross-sections of the unit cell. The structure with large rectangular micropillars (c) is optimized to achieve a high degree of circular polarization of photoluminescence in spontaneous mode $\rho_C \approx 60\%$ (“optimized” structure). The structure (d) is not optimized and demonstrates only a weak degree of circular polarization of photoluminescence in the spontaneous mode  $\rho_C \approx 4\%$.
    }
    \label{fig1:system}
\end{figure}

The purpose of this work is to investigate theoretically the properties of bi- and multistable transitions under resonant monochromatic pumping normal to the structure in a chiral semiconductor microcavity containing multiple quantum wells with cavity polaritons in the active region, as studied in papers~\cite{Demenev2016,Maksimov2022}. Self-consistent calculations are used to determine the polariton distribution across the quantum wells.
Previosly~\cite{Dmitrieva2023rus}, we solved this problem using the mean field approach without accounting for strong field variation in the different quantum wells.

Consider chiral semiconductor Bragg AlAs/AlGaAs microcavity, containing twelve GaAs quantum wells with cavity polaritons in the active region, as demonstrated schematically in fig.\,\ref{fig1:system}a.
For a detailed description of the structure, see, for example, ~\cite{Valenko2025rus}.
The initially achiral microcavity becomes chiral due to the photonic crystal layer with square lattice of rectangular micropillars, fabricated on its upper mirror (fig.\,\ref{fig1:system}b,c). This is because  the symmetry group of the structure (four-fold rotation axis $C_4$) does not contain mirror reflections.

Earlier, in our work ~\cite{Valenko2025rus}, we analyzed the symmetry of such microcavity resonant modes in vicinity of the $\Gamma$-point in the planar first Brillouin zone of the photonic crystal structure.
We demonstrated that, in agreement with the results of~\cite{Hopkins2016}, at the $\Gamma$-point (i. e., for photons, propagating normally to the microcavity plane) the main resonant mode of the microcavity is a polarization doublet.
In the basis of linear polarizations, both components of the doublet have orthogonal linear polarization in the microcavity active region and, due to chirality, elliptical polarizations in the far wave zone above the microcavity.
Furthermore, in accordance with the $C_4$ symmetry of the structure, the polarization ellipses of the doublet components are oriented normally to each other.
The latter allows us to write linear equations in the resonant approximation, linking the two-dimensional (2D) amplitudes $\vec{\mathcal{E}}=(\mathcal{E}_x,\mathcal{E}_y)$ of the doublet components --- that is, of the electric field in the region of quantum wells with polaritons --- with a linearly polarized coherent electromagnetic pumping field far from the system
$\vec{\mathcal{E}}_{\text{ext}} =(\mathcal{E}_{\text{ext},x},\mathcal{E}_{\text{ext},y})$ and the polaritons polarization in quantum wells $\vec{\mathcal{P}}=(\mathcal{P}_{x},\mathcal{P}_{y})$

\begin{gather}
    \left\{
        \begin{aligned}
            & 	(\omega - \omega_C) \mathcal{E}_{x}= a\mathcal{E}_{\text{ext},x} +
           ib\mathcal{E}_{\text{ext},y} + \beta \mathcal{P}_{x},\\
            &  (\omega - \omega_C) \mathcal{E}_{y}= -ib\mathcal{E}_{\text{ext},x} +
           a\mathcal{E}_{\text{ext},y} + \beta \mathcal{P}_{y}.
        \end{aligned}
    \right.
    \label{eq:nonlinear_PEsyst}
\end{gather}
Here, $\omega_C$ is the identical complex natural frequency of the resonant components of the doublet, $\beta$ is the exciton-photon coupling constant in the isotropic approximation,
and the real constants $a$ and $b$ describe the coupling of the linearly polarized field at the center of the structure with the linearly polarized coherent pump components at frequency $\omega$ normal to the structure.

Note that in real experiments with resonant pumping, a finite cross-sectional beam is used instead of a flat monochromatic wave. In addition, the photonic crystal itself, which is fabricated on the surface of the micro-resonator, has a finite size. This leads to the finiteness of the excitation region of the polariton modes in the reciprocal space around the $\Gamma$-point. However, when the pumping beam is normal to the structure, there are~\cite{Gavrilov2020rus} no significant changes in the bistability and multistability effects discussed below.

For convenience, we can now switch to the circularly polarized basis $\mathcal{E}_{\pm}=\frac{1}{\sqrt{2}}(\mathcal{E}_{x}\mp i \mathcal{E}_{y})$ (and similarly for
$\mathcal{E}_{\text{ext},\pm}$),  $\mathcal{P}_{\pm}=\frac{1}{\sqrt{2}}(\mathcal{P}_{x}\mp i \mathcal{P}_{y})$ of the  right (+) and left (-) circular polarizations
\begin{equation} \label{Eq:lin}
(\omega - \omega_C) \mathcal{E}_{\pm}= \alpha_{\pm} \mathcal{E}_{\text{ext},\pm} + \beta \mathcal{P}_{\pm},
\end{equation}
where $\alpha_{\pm}=a \mp b$.
To complete the description, following the approach developed in~\cite{Gippius2004,Gippius2007}, we add to the linear equation (\ref{Eq:lin}) the nonlinear Gross-Pitaevskii equation for the coupling of the effective electric field in the quantum well with the excitonic polarization in $\pm$ circular polarizations
\begin{equation} \label{Eq:nonlin}
(\omega - \omega_X) \mathcal{P}_{\pm} = A \mathcal{E}_{\pm} + F |\mathcal{P}_{\pm}|^2 \mathcal{P}_{\pm}.
\end{equation}
Here $\omega_X$ is the excitonic resonance frequency, $A$ is a constant, proportional to the excitonic oscillator strength, and the constant $F$ describes the efficiency of so-called blue shift of the resonant excitonic frequency due to repulsion of excitons with same spin projections, proportional to the excitonic concentration,
\begin{equation}\label{Eq:TildeOmega}
    \tilde{\omega}_{X,\pm} = \omega_X + F \abs{\M P_{\pm} }^2 .
\end{equation}
We neglected the weak attraction between excitons with opposite spin projections, following \cite{Gippius2007}.

Strictly speaking, the susceptibility of excitons in quantum wells is spatially non-local and depends on the form factor of electrons and holes $F(z) = \Psi_e(z)\ Psi_h(z)$, where $\Psi_{e(h)}(z)$ are the quantized wave functions of the transverse motion of electrons (holes) of the exciton in the quantum well~\cite{Keldysh1988,Gippius1994}, and has the form
\begin{multline}
    \mathcal{P}_{\pm}(z) = \frac{1}{\omega - \tilde{\omega}_{X,\pm }}\int \hat{A}(z,z') \mathcal{E}_{\pm}(z') dz', \\ \hat{A}(z,z')\propto F(z)F(z').
\end{multline}
However, when the microcavity is excited normally to the quantum wells the effective electric field is almost parallel to them. In this case, as demonstrated in~\cite{Gippius1994}, the system response does not depend on the form factor, and, in particular, we can assume that $\hat{A}(z,z')=A\delta(z-z'), A=\text{const}$ (local excitonic response approximation).
In what follows, we will use this approximation for exciton susceptibility, and to account for the blue shift in formula (\ref{Eq:TildeOmega}),
we will use the average value of exciton polarization across the quantum well, $\langle \abs{\M P_{\pm} }^2\rangle$.

Note that the resulting system of equations (\ref{Eq:lin},\ref{Eq:nonlin}) for a chiral micro-resonator with the $C_4$ symmetry group, firstly,
does not contain terms mixing opposite circular polarizations, and, secondly, all coefficients except $\alpha_{\pm}$ do not depend on the index $\pm$. The fact that $\alpha_{+} \neq \alpha_{-}$ for a chiral structure leads to the fact that in the spontaneous mode, the photoluminescence (PL) of a chiral microresonator is partially circularly polarized~\cite{Maksimov2014}. Using electrodynamic reciprocity~\cite{LLECM1984rus}, it can be shown that the degree of circular polarization of the PL of a chiral microresonator is equal to

\begin{equation}
    \rho_{C,PL} = \frac{I_+ - I_-}{I_+ + I_-} = \frac{\alpha_+^2 - \alpha_-^2}{\alpha_+^2 + \alpha_-^2},
\end{equation}
where $I_{\pm}$ is the PL intensity in $\pm$ circular polarization.

In the future, we will be interested in how the chiral structure responds to resonant pumping by a flat, monochromatic electromagnetic wave perpendicular to the structure. We will characterize the response of polaritons in quantum wells by their degree of circular polarization, defined as
\begin{equation}
    \rho_C = \frac{\abs{\M P_+}^2 - \abs{\M P_-}^2}{\abs{\M P_+}^2 + \abs{\M P_-}^2} .
\end{equation}

\section{Mean field approximation.}
To simplify the task, let us first assume that the polaritonic polarization is uniform in the plane of the wells and can be estimated as the average across all wells.
The relationship between external pumping and averaged polaritonic polarization can be calculated as follows.
First, we will neglect exciton nonlinearity. Using the Fourier modal method and the optical scattering matrix~\cite{Whittaker1999,Tikhodeev2002}, we will calculate the frequency dependence of the electric field spatial distribution in the structure, as described in paper~\cite{Demenev2016}. We will then determine all the parameters of the resonance approximation in equations (\ref{Eq:lin},\ref{Eq:nonlin}). The only parameter we will not determine is the nonlinearity coefficient $F$.
However, this coefficient only determines the characteristic scale of pump intensity $I_0= \frac c {8 \pi} \left|\frac A F \right| $ for observing bi- and multistability effects~\cite{Gippius2004,Gippius2007} without qualitative changes to system behavior. It can be shown that the order of magnitude of $I_0$ is $ \sim 0.4$ kW/cm$^2$.

\begin{figure}
	\centering
	\includegraphics[width = 0.5\textwidth]{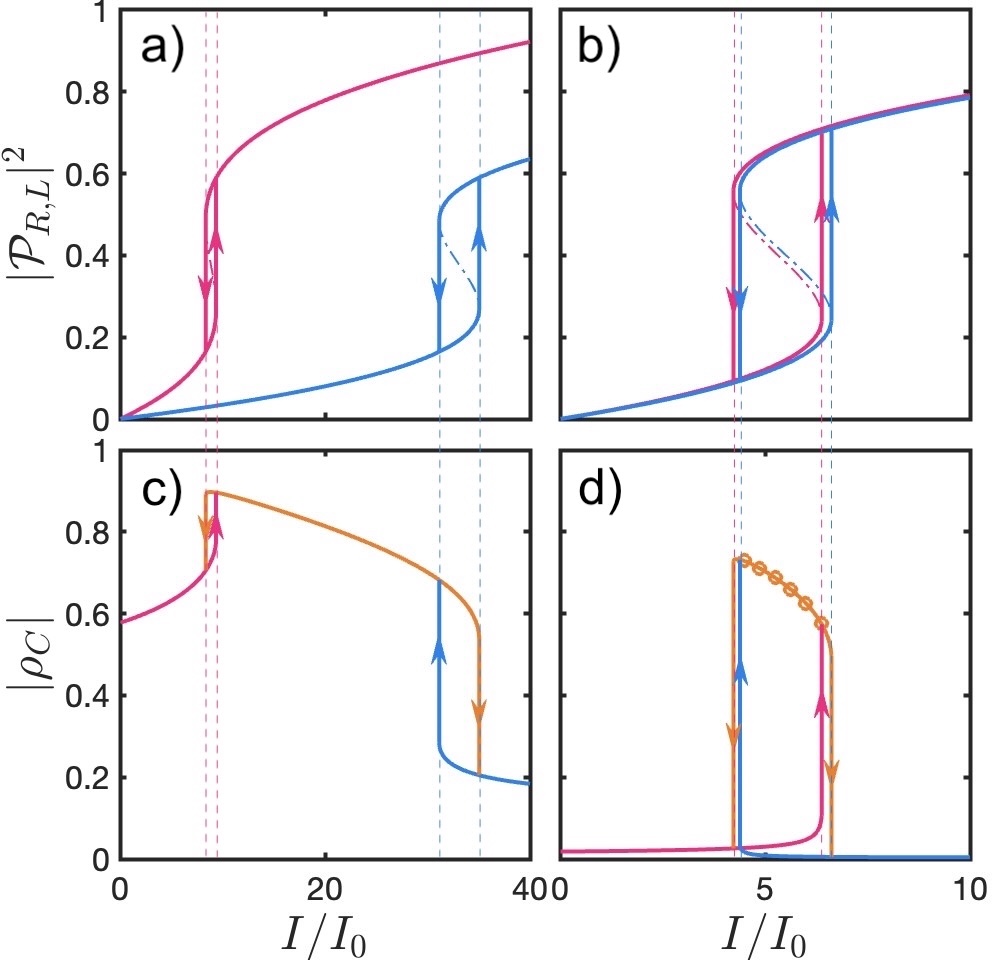}
    \caption{(a,b)
    The S-shaped dependence of the polaritonic intensity on the intensity of the external circularly polarized pumping in a chiral microresonator is shown at a fixed positive detuning of 0.4~meV.
   The solid red (blue) lines represent stable branches for predominant left (secondary right) circular polarization. The dashed lines show the unstable branches of the S-shaped dependencies.
   (c,d) The corresponding dependencies of the degree of circular polarization of polaritons on the intensity of linearly polarized pumping, calculated with the same frequency detuning.  Panels (a) and (c) refer to the optimized structure, while panels (b) and (d) refer to the non-optimized structure. The arrows in all panels indicate the directions of transitions during bistable jumps, the threshold intensities of which are marked by red (for left circular polarization) and blue (for right circular polarization) vertical dashed lines. The yellow line with round markers in panel d) indicates the multistable branch, which is absent in transitions in the optimized structure (see explanation in the text).}
    \label{fig2:S&rho}
\end{figure}

By excluding the electric field from the system of equations~(\ref{Eq:lin},\ref{Eq:nonlin}), we obtain a cubic equation for the amplitude of the average polaritonic polarization $\M P_{\pm}$, which describes our system with account for excitonic nonlinearity
\begin{multline}
	\M P_{\pm} \left[(\omega - \omega_C)(\omega - \omega_X - F \abs{\M P_{\pm}}^2) - \Omega_R^2 \right]
	= \\ =
	{A \alpha_{\pm}}  \mathcal{E}_{\text{ext} \pm}.
	\label{eq:cubic_for_P}
\end{multline}
As is well known, the solution to this equation for steady-state dependencies, $\abs{\M P_{\pm}(I)}^2, I = \abs{\M E_{\text{ext},\pm})}^2$, for positive pump frequency detuning from the exciton resonance frequency and a certain range of pump intensity takes the form of an S-shaped curve~\cite{Gippius2004,Gippius2007}. This curve provides the possibility of a bistable response (Fig.~\ref{fig2:S&rho}a,b).

The magnitude of detuning here and below is chosen to be in the region of maximum hysteresis effects and is equal to 0.4~meV. With further increase in the detuning, bistability disappears~\cite{Gavrilov2020rus,Barulin2024}.
As equation (\ref{eq:cubic_for_P}) clearly shows, the dependencies of the different components of circular polarization only differ in scale along the horizontal intensity axis.

Thus, the threshold values of the intensity of bistable transitions up and down in the intensity of polaritons with different spin projections (vertical dashed lines in Fig.~\ref{fig2:S&rho}) are related by the following equation:
\begin{equation}
    I_{\text{th},+}=\frac{\alpha_-^2}{\alpha_+^2}I_{\text{th},-}.
\end{equation}
This equation shows that the threshold values differ from each other.

In the model chiral structures described earlier, left circular polarization is predominant ($\alpha_-> \alpha_+$).
As a result, the bistability thresholds
of the right circular component are higher than those of the left component by a factor of $\alpha_-^2/\alpha_+^2 \sim 4 $ for the optimized structure and only $\sim 1.08$ times for the unoptimized structure.
	
\begin{figure}
	\centering
	\includegraphics[width = 0.3\textwidth]{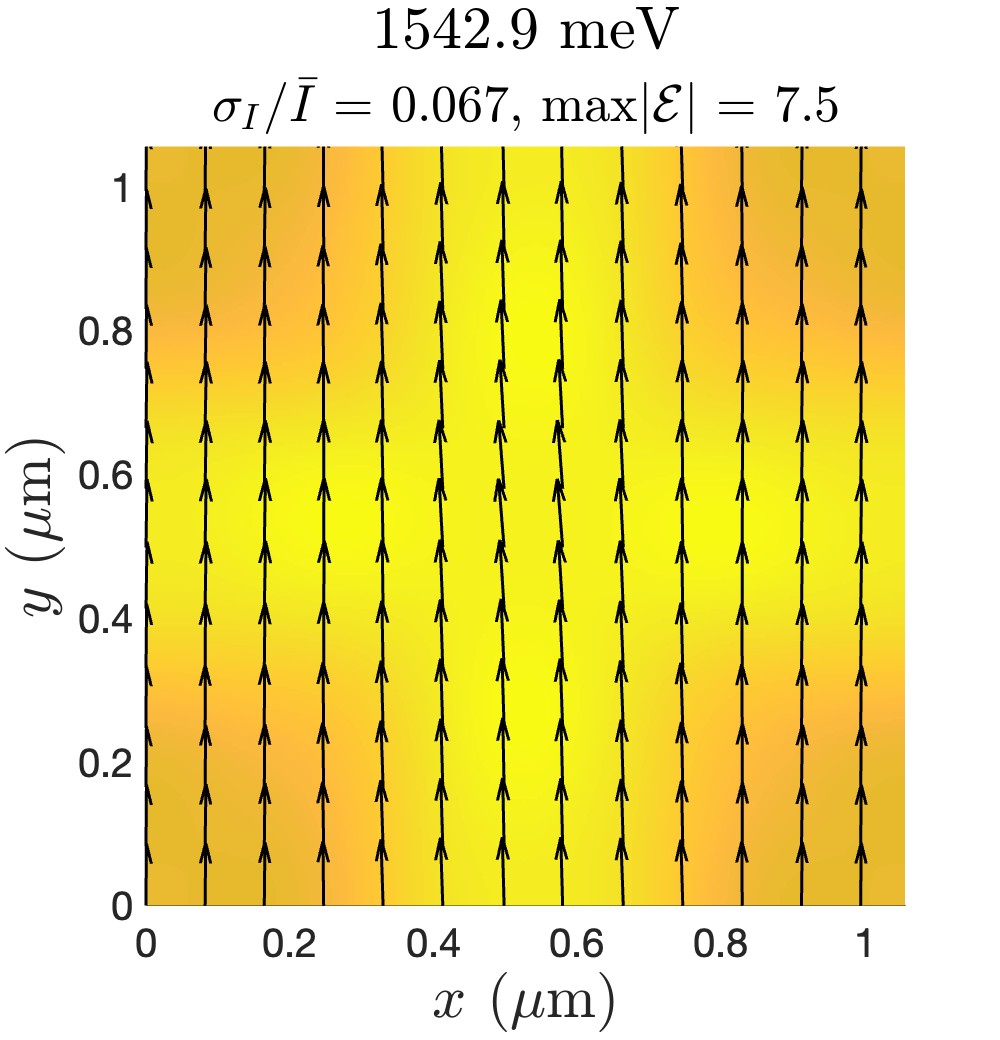}		
    \includegraphics[width = 0.3\textwidth]{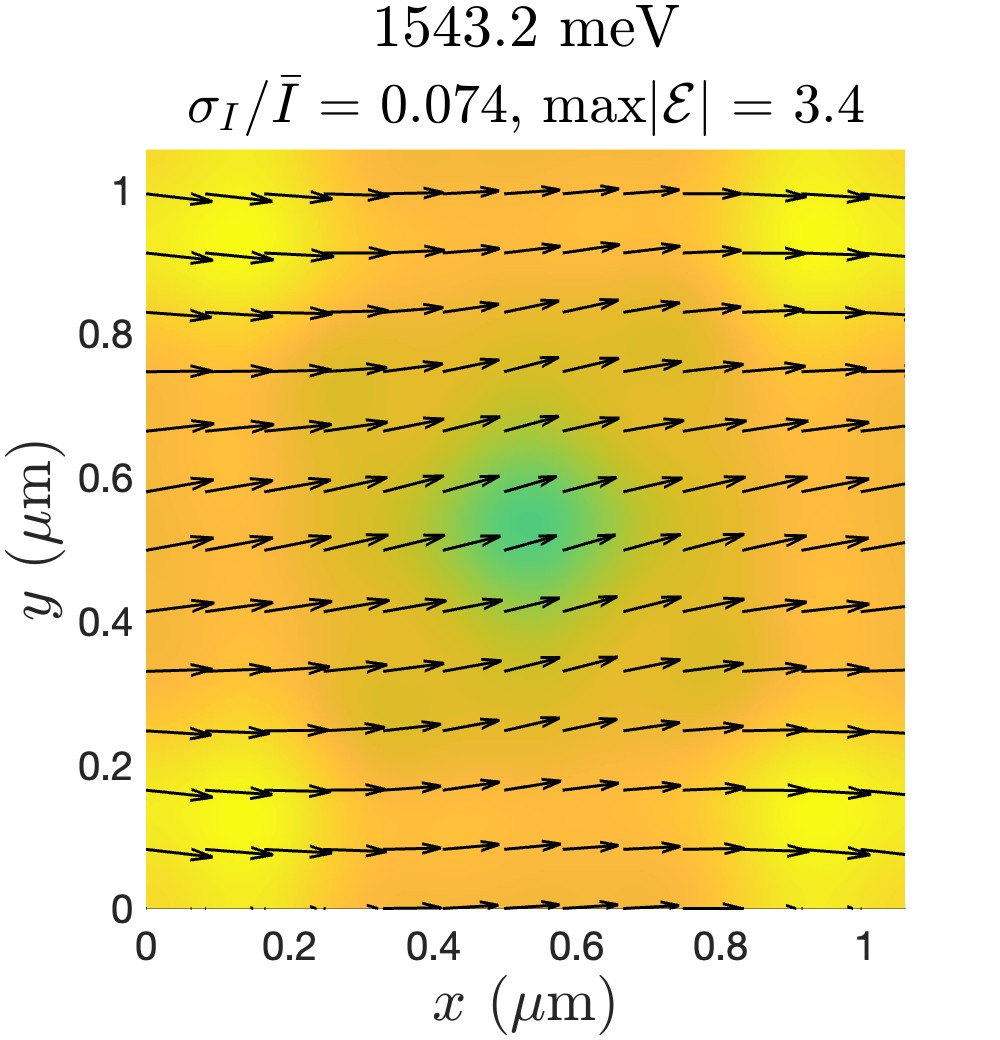}	
    \includegraphics[width = 0.33\textwidth]{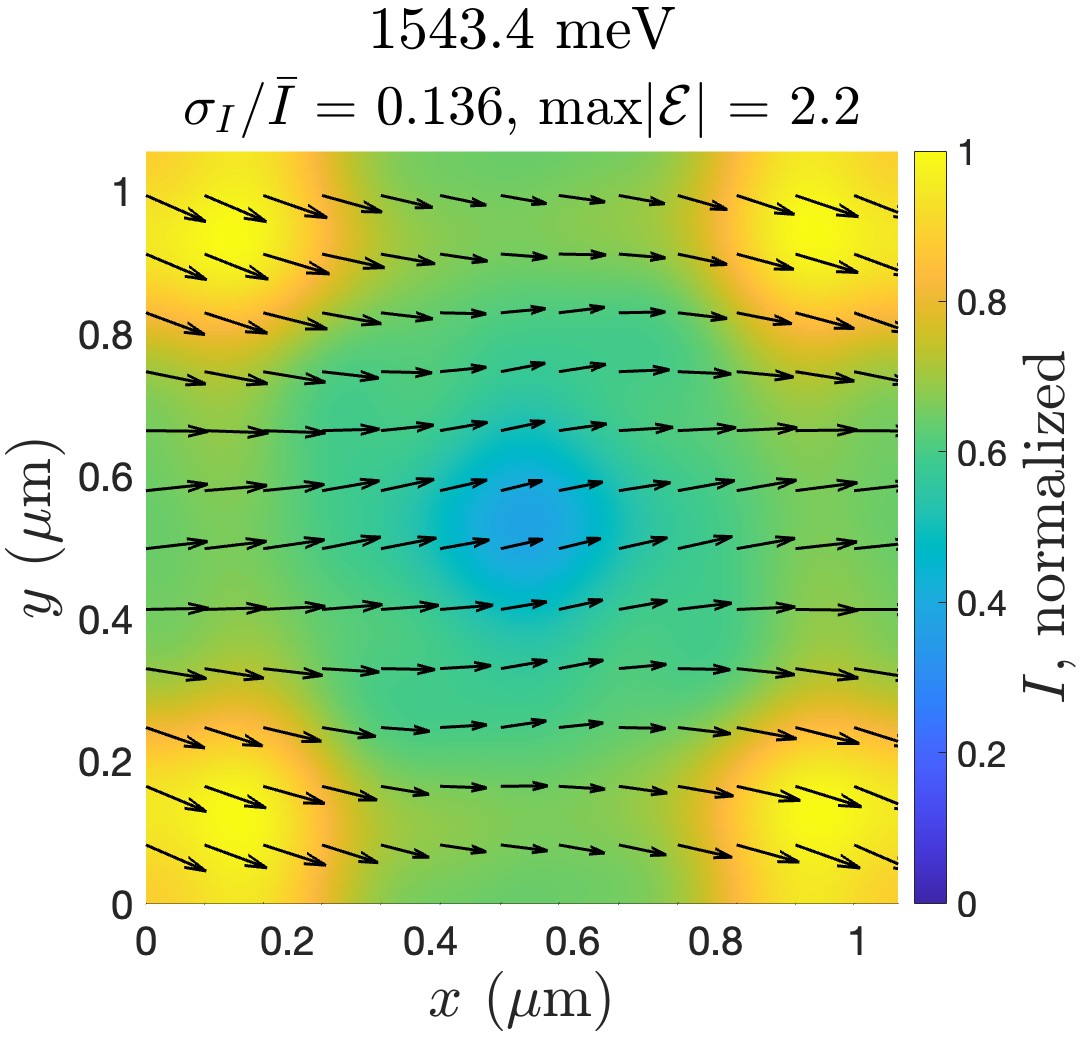}	
	\caption{The normalized distribution of the electric field in the plane of the quantum well of the non-optimized structure in the case of circularly polarized pumping of weak intensity (in linear mode) normal to the structure. One elementary cell of the photonic crystal is shown. The colored background on all panels shows the distribution of the normalized electric field intensity (see the color scale on the right). The pumping frequency, the maximum electric field $\mathrm{max}|\mathcal{E}|$ in units of the pumping field, and the normalized root-mean-square deviation of the intensity distribution $\sigma_I/\overline{I}$ are shown in the panel captions. The resonance frequency of the lower polariton at the $\Gamma$-point is 1542.9~meV.}
    \label{fig3:field_xy}
\end{figure}

In the case of linearly polarized pumping, the intensities of the left- and right-polarized circular components of the pump are equal.
As pump intensity increases, the bistability threshold of the predominantly left-polarized component is reached first.
 Therefore, the intensity of the left-polarized exciton component jumps from the lower to the upper branch of the S-shaped contour. Meanwhile, the right-polarized components remain on the lower branch because the corresponding threshold is higher in intensity.
During this transition, the degree of circular polarization increases significantly. In the optimized structure, it rises from about 0.7 to 0.9. In the non-optimized structure, however, it rises even more dramatically, from 0.05 to 0.7.

As pumping intensity increases further, the bistability threshold for the secondary circular polarization (right in this case) is also reached. Consequently, the intensity of the right-polarized polaritons jumps to the upper branch, and the circular polarization degree $\rho_C$ decreases abruptly.

In the case of a non-optimized system, there is another branch (shown in Fig. \ref{fig2:S&rho}d as a yellow line with round markers) that cannot be reached with the monotonic increase in pumping intensity considered above. However, if after overcoming the first threshold the pumping intensity is reduced without reaching the
second threshold, the system remains in a state of “left polarization on the upper branch, right polarization on the lower branch” of its S-circuits until the intensity
decreases beyond the threshold of the reverse transition of left polarization to the lower branch.

\begin{figure}
    \centering
	a)	\includegraphics[width = 0.4\textwidth]{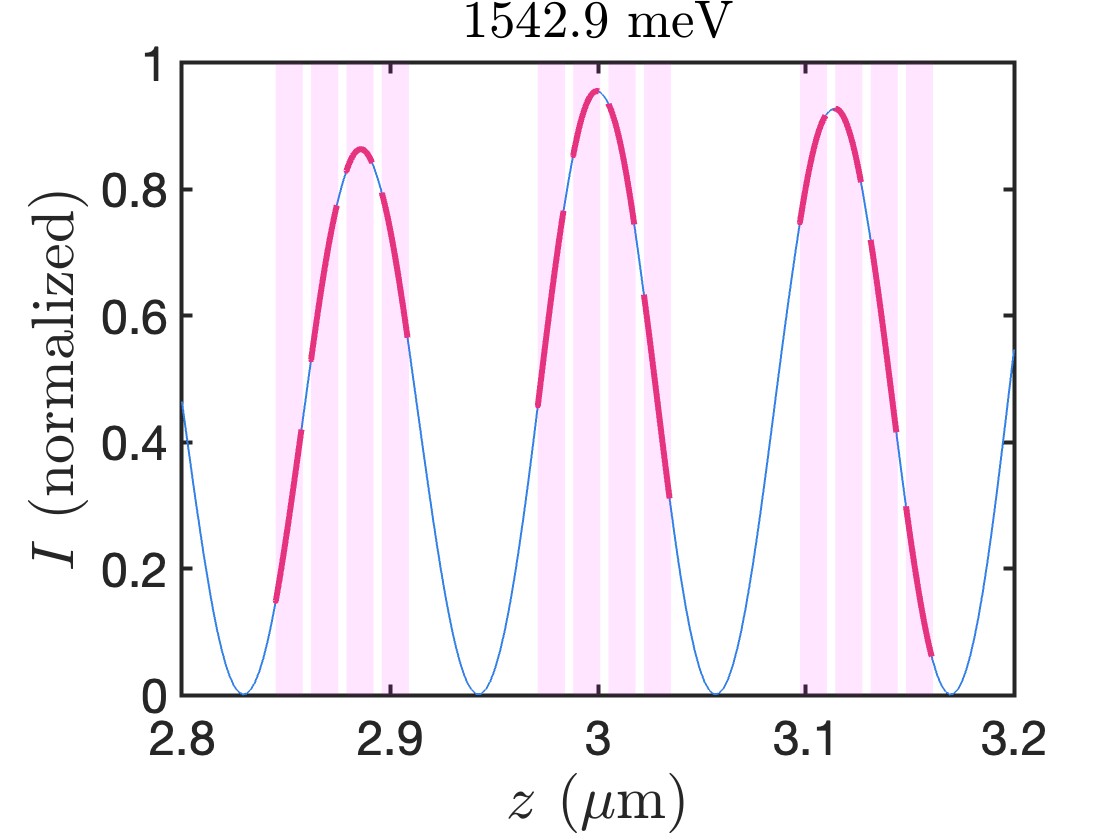}	\\	
    b) \includegraphics[width = 0.4\textwidth]{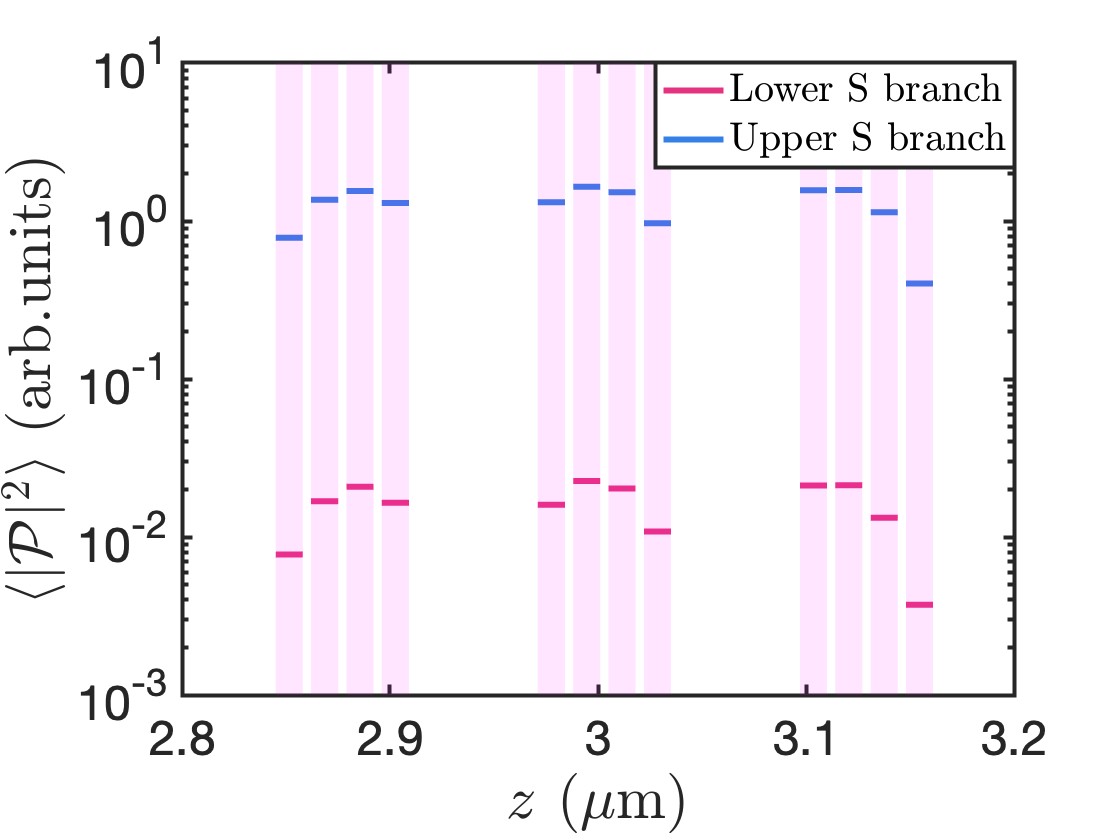}	

    \caption{(a) The electric field intensity $|\mathcal{E}|^2$ distribution in the quantum well region (vertical red stripes) in the $z$ direction in the case of weak
    	excitation (linear regime).  (b) Distribution of the average intensity of excitons in the quantum wells of the structure in the $z$ direction in the nonlinear mode on the upper (blue stripes) and lower (red) branches of the bistable S-circuit.}
    \label{fig4:field_z}
\end{figure}

Note that in an achiral microresonator, i.e., before a chiral photonic crystal is fabricated on it, $\alpha_+ =\alpha_-$. However, the difference in the right-hand sides of equation (8) for different circular polarizations required for multistable transitions in Fig. 2 can be ensured by selecting the appropriate elliptical polarization of the pump. This was demonstrated in the work~\cite{Gavrilov2013}.

\section{Self-consistent accounting for the heterogeneity of the spatial distribution of exciton polarization.}

The above results were obtained assuming that polarization is distributed almost uniformly across quantum wells. Now let us discuss a method for accounting for its heterogeneity.
As the calculation shows, the field distribution in the plane of quantum wells is approximately uniform near resonance. If the frequency deviates from resonance, the field modulation gradually increases, though only within relatively small limits (Fig.~\ref{fig3:field_xy}).
The most significant variations in the amplitude of the electric field are along the axis normal to the structure, i.e., in different quantum wells.
The typical electric field intensity distribution in various quantum wells in linear regime (for weak circularly polarized pumping) is shown in fig.~\ref{fig4:field_z}a also using the example of a non-optimized structure.

Let us consider how to calculate self-consistently the polariton density distribution along the $z$ axis, perpendicular to the structure, taking into account the nonlinear effect.	
The Fourier-modal method and the scattering matrix formalism allow us to find
the linear approximation of the field distribution $\mathbf{E}_{\omega}(x,y,z)$ in the system for given values of frequency and complex vector amplitude of the incident field. For this calculation, it is necessary to specify the distribution of dielectric permittivity in the system.

In our case, the subject of consideration is a nonlinear effect, as a result of which, in a strong electric field in the system, due to the blue shift of the exciton frequency
(\ref{Eq:TildeOmega}) $\Delta \omega_X = \tilde{\omega}_X - \omega_X$, we can talk about a change in the dielectric permeability of the quantum well material:
$\tilde{\varepsilon}_{QW}(\omega) = \varepsilon_{QW}(\omega-\Delta \omega_X)$.

In this case, if the values of $\Delta \omega_X$ are known for each layer for the given parameters of the incident wave, and the modified values of the dielectric permittivity in the quantum wells $\tilde{\varepsilon}_{QW}(\omega)$ are specified for them, then when calculating the field distribution in the system using the scattering matrix method, the correct field distribution in the system will be obtained, consistent with the distribution of the polaritonic density and the corresponding blue shift in the quantum wells.
At the same time, although the scattering matrix method is intended for analyzing the behavior of the system in the linear case, nonlinearity is taken into account by the modified dielectric permittivities of layers with quantum wells.

\begin{figure}
	\centering
	\includegraphics[width = 0.43\textwidth]{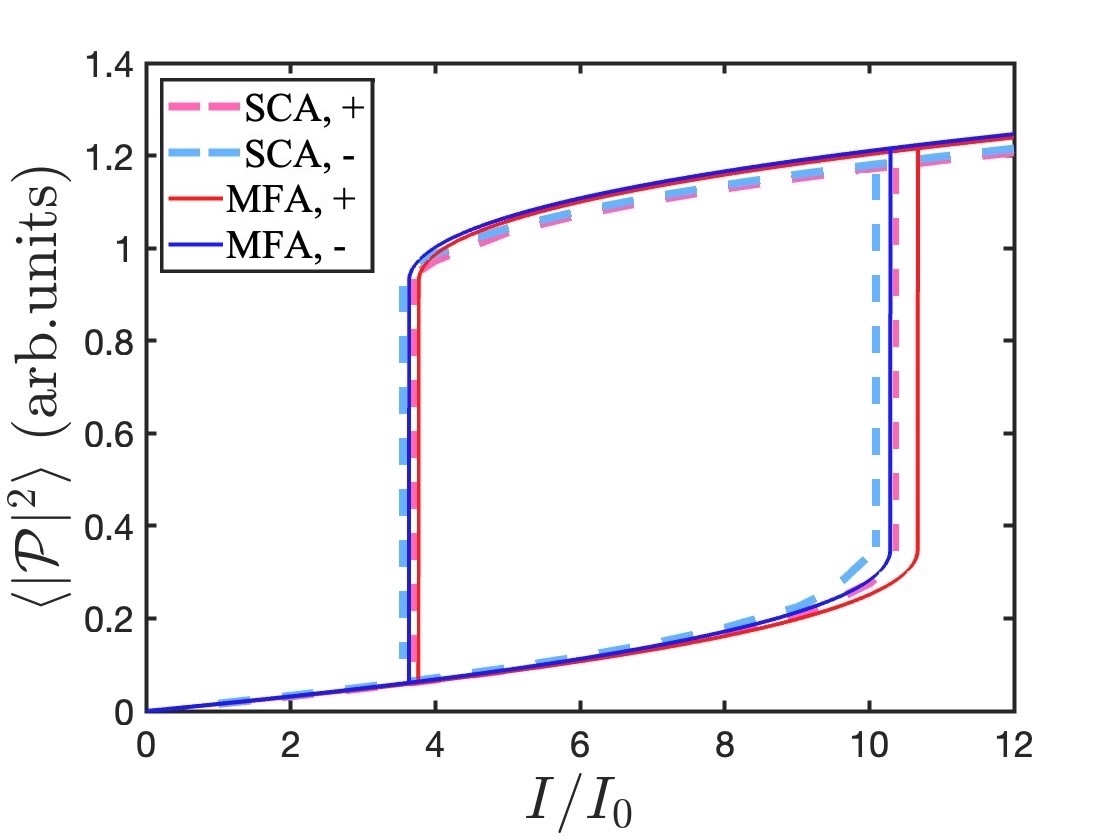}	
  	\caption{Comparison of the dependencies of the squared average polaritonic polarization amplitude on the pump intensity, calculated within the mean field approximation (MFA) using a cubic equation, and within the self-consistent approximation (SCA) approach, which takes into account the variation of the field distribution in different quantum wells.}
	\label{fig5:bistability}
\end{figure}

The following method was used to find the field distribution in the system at fixed parameters of the incident wave.
First, a certain initial distribution of exciton density was specified, and based on it, the values of dielectric permittivity in all layers of the system were determined (taking into account the blue shift of the excitonic frequency).
Next, based on the known distribution of dielectric permittivity, the electric field at each point of the system was calculated using the scattering matrix method.
Based on the obtained electric field distribution, the polarization distribution of the quantum wells at each point was determined, and the average value of the square of the modulus of this polarization, and thus the polaritonic density, was found within each of the layers.
The obtained values of the exciton density were compared with the initial ones, and the discrepancy between them was minimized iteratively using Newton's method.
The results of this analysis are shown in Fig.~\ref{fig4:field_z}b for the self-consistent distribution of polariton intensity across all 12 quantum wells on the lower (upper) contour of the S-shaped bistable contour by red (blue) horizontal lines.
Self-consistent analysis was performed for the case of an unoptimized structure and preferential (left) circular polarization of the pump with a detuning from the polariton frequency at a low pump of 0.6~meV normal to the structure.
In Fig.~\ref{fig5:bistability}, the self-consistent bistability contours calculated for $\pm$ polarized pumping are shown as dashed lines, compared to the contours calculated in the mean-field approximation
and shown as solid lines. This comparison shows that self-consistent consideration of the inhomogeneity in the distribution of polaritons in the quantum wells of the system does not lead to qualitative differences from the mean-field approximation.

\section{Conclusion}
In conclusion, we have demonstrated that the nonlinearity of cavity polaritons and the bistability of their response to circularly polarized pumping in a chiral microcavity lead to multistability of the response to linearly polarized pumping.
At the same time, the degree of circular polarization of polaritons can reach significantly higher values than in their photoluminescence in spontaneous mode.
For example, a non-optimized structure with $\rho_{C,PL} \sim 4\%$ in the spontaneous mode, taking into account nonlinearity and resonant pumping, can reach a degree of circular polarization of 60-80\% even with linearly polarized pumping.
It is shown that a self-consistent calculation of the distribution of polariton density across quantum wells in the structure does not lead to a qualitative difference from the mean-field approximation.

\pagebreak[4]

% \textbf{Благодарности.}
The authors are grateful to A. A. Bogdanov,  V. D. Kulakovsky, and Thomas Weiss for the discussion.

%Авторы благодарны А. А. Богданову,  В. Д. Кулаковскому и Томасу Вайссу (T.\,Weiss) за обсуждение.
% \textbf{Финансирование работы.}
This research was supported by a grant from the Russian Science Foundation (\textnumero 22-12-00351-\textcyr{П}).
%Исследование выполнено за счёт гранта Российского научного фонда № 22-12-00351-П.

% \textbf{Конфликт интересов.}
% Авторы данной работы заявляют, что у них нет конфликта интересов.


\begin{thebibliography}{10}

\bibitem{Dmitrieva2026}  O. A. Dmitrieva, N. A. Gippius, and S. G. Tikhodeev, Zh. Exp. Teor. Fiz. {\bf 169}, 181 (2026) (in Russian). %

\bibitem{Konishi2011}  K.\,Konishi, M.\,Nomura, N.\,Kumagai, S.\,Iwamoto, Y.\,Arakawa, M.\,Kuwata-Gonokami, Phys. Rev. Lett. {\bf 106}, 057402 (2011). %1

\bibitem{Maksimov2014} A. A. Maksimov, I. I. Tartakovskii, E. V. Filatov, S. V. Lobanov, N. A. Gippius, S. G. Tikhodeev, C. Schneider, M. Kamp, S. Maier, S. H\:ofling, and V. D. Kulakovskii, Phys. Rev. B {\bf 89}, 045316 (2014).

\bibitem{Lobanov2015}	 S.\,V.\,Lobanov, T.\,Weiss, N.\,A.\,Gippius, S.\,G.\,Tikhodeev, V.\,D. Kulakovskii, K.\,Konishi,  M.\,Kuwata-Gonokami,  Opt. Letters {\bf 40}, 1528 (2015). %3

\bibitem{Demenev2016} A.\,A.\,Demenev, V.\,D.\,Kulakovskii, C.\,Schneider, S.\,Brodbeck, M.\,Kamp,  S.\,H\"{o}fling, S.\,V.\,Lobanov, T.\,Weiss, N.\,A.\,Gippius, S.\,G.\,Tikhodeev, Appl. Phys. Lett. {\bf 109}, 171106 (2016). %4

\bibitem{Tanaka2020}  K.\,Tanaka, D.\,Arslan, S.\,Fasold, M.\,Steinert, J.\,Sautter, M.\,Falkner, T.\,Pertsch, M.\,Decker, I.\,Staude, ACS Nano {\bf 14}, 15926 (2020). %5

\bibitem{Qu2021}  D.\,Qu, M.\,Archimi, A.\,Camposeo. D.\,Pisignano, E.\,Zussman, ACS Nano {\bf 15},  8753 (2021). %6

\bibitem{Gao2021} X.\,Gao, Y.\,Xu, J.\,Huang, Z.\,Hu, W.\,Zhu, X.\,Yi, L.\,Wang,  Opt. Lett. {\bf 46}, 2666 (2021). %7

\bibitem{Fong2021} C.\,F. Fong, Y.\,Ota, Y.\,Arakawa, S.\,Iwamoto, Y.\,K.\,Kato,  Phys. Rev. Res.  {\bf 3},   043096 (2021). %8

\bibitem{Zhang2022} X.\,Zhang , Y.\,Liu, J.\,Han, Y.\,Kivshar, Q.\,Song, Science  {\bf 377}, 1215 (2022). %9

\bibitem{Maksimov2022} A.\,A.\,Maksimov, E.\,V.\,Filatov, I.\,I.\,Tartakovskii, V.\,D.\,Kulakovskii, S.\,G.\,Tikhodeev, C.\,Schneider,  S.\,H\"{o}fling, Phys. Rev. Applied {\bf 17}, L021001 (2022). %10

\bibitem{Toftul2024} I. Toftul, P. Tonkae, K. Koshelev, F. Lai, Q. Song, M. Gorkunov, Y. Kivshar, Phys. Rev. Lett. {\bf 133}, 216901 (2024). %11

\bibitem{Tsai2024} J.-T.\,Tsai, C.-T.\,Hsieh, C.-C.\,Cheng, M.-H.\,Shih,S.-W.\,Chang, Laser \& Phot. Rev. {\bf 18}, 2300574 (2024). %12

\bibitem{Takahashi2025} S.\,Takahashi, Y.\,Kinuta, S.\,Ito, H.\,Onishi, K.\,Yamashita, J.\,Tatebayashi, S.\,Iwamoto, Y.\,Arakawa, Appl. Phys. Lett. {\bf 126}, 081108 (2025). %13

\bibitem{Gromyko2025} D.\,Gromyko, J.\,S.\,Loh, J.\,Feng, C.-W.\,Qiu, L.\,Wu, Phys. Rev. Lett. {\bf 134}, 023804  (2025). %14

\bibitem{Valenko2025rus} N. V. Valenko, N. A. Gippius,  V. D. Kulakovskii,  S. G. Tikhodeev,  JETP Letters {\bf 121}, 829 (2025)  %15

\bibitem{Maksimov2022arus}  A. A. Maksimov, E. V. Filatov, I. I. Tartakovskii, JETP Letters {\bf 116}, 500 (2022). %16	

\bibitem{Baas2004} A. Baas, J. Ph. Karr, H. Eleuch, and E. Giacobino, Phys. Rev. A {\bf 69}, 023809 (2004). %17

\bibitem{Gippius2004} N. A. Gippius, S. G. Tikhodeev, V. D. Kulakovskii, D. N. Krizhanovskii, and A. I. Tartakovskii, Europhys. Lett. {\bf 67}, 997 (2004). %18
		
\bibitem{Gippius2007} N. A. Gippius, I. A. Shelykh, D. D. Solnyshkov, S. S. Gavrilov, Y. G. Rubo, A. V. Kavokin, S. G. Tikhodeev, and G. Malpuech, Phys. Rev. Lett. {\bf 98}, 236401 (2007). %19

\bibitem{Sarkar2010} D. Sarkar, S. S. Gavrilov, M. Sich, J. H. Quilter, R. A. Bradley, N. A. Gippius, K. Guda, V. D. Kulakovskii, M. S. Skolnick, and D. N. Krizhanovskii, Phys. Rev. Lett. ,  {\bf 105} ,   216402 (2010). %20

\bibitem{Gavrilov2013} S. S. Gavrilov, A. V. Sekretenko, S. I. Novikov, C. Schneider, S. H\:ofling, M. Kamp, A. Forchel, and V. D. Kulakovskii,  Appl. Phys. Lett. {\bf 102}, 011104 (2013).

\bibitem{Brichkin2015} A. S. Brichkin, S. G. Tikhodeev, S. S. Gavrilov, N. A. Gippius, S.I. Novikov, A. V. Larionov, C. Schneider, M. Kamp, S. H\:ofling, and V. D. Kulakovskii, Phys. Rev. B {\bf 91}, 125155 (2015). %21

\bibitem{Gavrilov2020rus} S. S. Gavrilov, Physics--Uspekhi {\bf 63},  123 (2020). %22

\bibitem{Dmitrieva2023rus} O. A. Dmitrieva, N. A. Gippius,  S. G. Tikhodeev, Doklady Physics {\bf 68}, 144 (2023) %23

\bibitem{Hopkins2016} B. Hopkins, A. N. Poddubny, A. E. Miroshnichenko, and Y. S. Kivshar, Laser Photonics Rev. {\bf 10}, 137 (2016). %24


\bibitem{Keldysh1988} L. V. Keldysh, Superlattices \& Microstructures  {\bf 4/5}, 637 (1988). %26->25

\bibitem{Gippius1994} N. A. Gippius, S. G. Tikhodeev and L. V. Keldysh, Superlattices \& Microstructures  {\bf 15}, 479 (1994). %27->26

\bibitem{LLECM1984rus} L. D. Landau and E. M. Lifshitz, Electrodynamics of Continuous Media. Volume 8 in Course of Theoretical Physics. Second Edition (Elsevier Ltd. All, 1984), p. 309 %25->27
		
\bibitem{Whittaker1999} D. M. Whittaker and I. S. Culshaw, Phys. Rev. B {\bf 60}, 2610 (1999). %28
		
\bibitem{Tikhodeev2002} S. G. Tikhodeev, A. L. Yablonskii, E. A. Muljarov, N. A. Gippius, and Teruya Ishihara, Phys. Rev. B {\bf 66}, 045102 (2002). %29

\bibitem{Barulin2024} A. Barulin, O. Pashina, D. Riabov, O. Sergaeva, Z. Sadrieva, A. Shcherbakov, V. Rutckaia, J. Schilling, A. Bogdanov, I. Sinev, A. Chernov, M. Petrov, Laser Photonics Rev. {\bf 18}, 2301399 (2024).
 %30

\end{thebibliography}
\end{document}